\documentstyle[preprint,pra,aps]{revtex}

\begin{document}
\draft

\title{\Large \bf
       A light-fronts approach to electron-positron pair production
       in ultra-relativistic heavy-ion collisions }
\author{B. Segev and J. C. Wells\thanks{
New address: Oak Ridge National Laboratory, 
             P.O.\ Box 2008,
             Oak Ridge, TN 37831-6203}}
\address{
       Institute for Theoretical Atomic and Molecular Physics, \\
       Harvard-Smithsonian Center for Astrophysics, \\
       60 Garden Street, Cambridge, MA 02138}

\date{\today}

\maketitle
{\begin{abstract}
We perform a gauge-transformation on the time-dependent
Dirac equation describing the evolution of an electron
in a heavy-ion collision to remove the explicit dependence
on the long-range part of the interaction.
We solve, in an ultra-relativistic limit, the 
gauged-transformed Dirac equation 
using light-front variables and a light-fronts representation,
obtaining non-perturbative results for the 
free pair-creation 
amplitudes in the collider frame.
Our result reproduces the result of second-order 
perturbation theory in the small charge limit
while non-perturbative effects arise 
for realistic charges of the ions. 
\end{abstract}


\newpage

\section{Introduction}

Electron-positron pair production from the vacuum in strong electromagnetic 
fields is a fundamental prediction of 
QED\cite{IZ,greiner,eichler}. 
In heavy-ion collisions at energies near the Coulomb barrier,
quasi-bound molecular states are formed with binding energies 
which dive into the negative-energy continuum,
resulting in a resonance which decays into an electron-positron 
pair\cite{greiner,Ge70,Ma74,Mu73,Ra76,Li77}.
In contrast, for {\it ultra-relativistic} heavy-ion collisions
at peripheral impact parameters,
the ions execute straight-line trajectories
and reside near each other for only a very short time.
The high charge of the individual ions and
the strong Lorentz contraction combine to produce fields sufficient
for electromagnetic pair production through a 
qualitatively different process.
Large cross sections for electro-magnetic pair 
production in these collisions were theoretically 
predicted\cite{Ei90} and experimentally 
observed \cite{VD92,BG93,Ba94b,BG94,VD94,VD95,DE95,CB97}.

In lowest-order perturbation theory, the amplitudes for pair 
production in heavy-ion collisions are calculated from 
two-photon exchange diagrams\cite{BS89a}. 
The quantum field theoretical treatment of this process was 
reduced to a classical-field approach\cite{WB90}.
Experimental observations of free pair production in the
energy range around ten GeV per nucleon 
(in the collider reference frame)
are in agreement with second-order perturbation 
theory\cite{VD92,Ba94b,VD94,VD95,DE95}.
For lower energies of a few GeV per nucleon (collider frame), 
experimental results for free and bound-free pair production
show deviations from the predictions obtained from
two-photon exchange diagrams\cite{BG93,BG94}.
This is likely due to two-center Coulomb effects\cite{Ei95,MB96}. 
In the near future, larger ultra-relativistic energies 
above one hundred GeV per nucleon (collider frame) will be 
available.
New non-perturbative effects may 
become important at colliding-beam accelerators such 
as the Relativistic Heavy-ion Collider (RHIC) at Brookhaven 
and possibly the Large Hadron Collider (LHC) at CERN.
These non-perturbative effects are the subject of our present
work.

Previous theoretical works on high-energy 
non-perturbative effects considered
unitarity violation in lowest-order 
perturbation theory, multiple-pair production,
and corrections to production cross 
sections\cite{Ba90b,RW91,BRW,BG92,Io94,HT95,GW95,Ba95,BRW96,AH97,Baltz97}.
An open question, crucial for the beam stability, is whether 
the non-perturbative effects will enhance or reduce 
the cross section for bound-free production, i.e.\
production with capture\cite{BRW96,Baltz97}.

In a non-perturbative treatment, starting from the 
QED Lagrange density operator,
the Euler-Lagrange equations of motion for the quantum 
fields are equivalent, under physical assumptions, to 
the one-particle Dirac equation interacting with classical,
electromagnetic fields \cite{WO92}. 
Calculations of probabilities and correlations
can then be reduced 
to solving the two-center time-dependent Dirac equation,
which describes the dynamics of an electron in the classical
field of two relativistically moving charges.

In the ultra-relativistic limit, the ions
are practically moving at the speed of light.
The classical electromagnetic field of a massless charged particle
was studied in Ref.\ \cite{Ja92}.
It can be described by pure gauge potentials, with different 
gauges in different regions of space-time. 
The eikonal approximation was then reproduced from an 
exact solution of a quantum-mechanical equation in this field.
A similar approach was recently used in target-frame 
calculations of bound-free pair production,
where a gauge transformation was used to remove the
long-range Coulomb effects \cite{Ba95,Baltz97}.

The use of gauge transformations is fundemental 
to these calculations.  
The term gauge transformation, 
as used here, is not to be confused with a gauge-symmetry 
transformation in which both the wave function and the fields 
are transformed so as to keep the equation of motion invariant. 
Here, as well as in Refs.~\cite{BRW,Ru90,Ru93}, for example, 
a space-time dependent phase is used to transform {\it either}
the wave function {\it or} the fields in order to obtain a different
equation of motion.
The connection between the solution of the original problem and 
the gauge-transformed problem depends on the asymptotic
(infinite time) behavior of the gauge function employed;
i.e.\ on the induced changes to the initial and 
final states\cite{BRW,Ru90,Ru93}.
Gauge transformations should also be applied with care
when used in calculations employing truncated basis 
sets\cite{Ru90,Ru93,EB96}. 

In this work, a gauge transformation is used 
to solve the two-center Dirac equation 
describing an electron during a relativistic 
heavy-ion collision.
A closed form expression for the pair-production 
amplitudes in the ultra-relativistic limit is found.
First, in section II, the 
ultra-relativistic limit for the two-center Dirac equation 
is obtained and discussed. In section III, 
exact and asymptotic relations between the 
physical amplitudes and the
gauged transformed ones are established. In section IV, the 
{\it light-fronts representation} is introduced and the 
foundations are laid for the construction of the exact 
solution in section V,
where the perturbative limit is considered as well. 
The physical contents of our results 
and an outlook for future applications are 
finally considered in section VI.
Details of some derivations are given in appendices.

\section{An ultra-relativistic limit to the two-center  
time-dependent Dirac equation} 

Consider a collision between two ions with 
charges $Z_A$ and $Z_B$ and velocities $\beta \hat{z}$ and 
$-\beta \hat{z}$, respectively, moving parallel to each other 
at an impact parameter of $2 \vec{b}$ (see Fig.\ \ref{coll_fig}).
An external-field approach to the influence of these ions on 
the vacuum is appropriate for 
peripheral impact parameters (i.e.\ no nuclear interactions),
heavy ions, and ultra-relativistic energies, 
when to a very good approximation, the ions 
continue intact on their parallel, 
straight-line trajectories. 
The two-center Dirac equation for an electron in the 
time-dependent external field of these ions is given by:
   \begin{equation}
i\frac{\partial}{\partial t} |\Phi(\vec{r},t)\rangle =
\left[ \hat{H}_0 + \hat{H}_A(t) + \hat{H}_B(t) \right] 
|\Phi(\vec{r},t)\rangle \label{dirac},
\end{equation}
where $|\Phi(\vec{r},t)\rangle$ is the Dirac spinor wave 
function of the 
electron, $\hat{H}_0$ is the free Dirac Hamiltonian and 
$\hat{H}_A(t)$ and $\hat{H}_B(t)$ are each the interaction 
with one ion,
   \begin{eqnarray}
&&\hat{H}_0 \equiv -i \check{\alpha}\cdot\vec{\nabla} + 
\check{\gamma^0} \label{H_0},\\
&&\hat{H}_A(t)\equiv ({\rm I}_4 - \beta\check{\alpha}_z)
\frac{-Z_A \alpha}
{\sqrt{(\vec{r}_{\perp}-\vec{b})^2/\gamma^2+(z-\beta t)^2}} 
\label{H_A},\\
&&\hat{H}_B(t)\equiv ({\rm I}_4+\beta\check{\alpha}_z)
\frac{-Z_B \alpha}
{\sqrt{(\vec{r}_{\perp}+\vec{b})^2/\gamma^2+(z+\beta t)^2}}
\label{H_B}.
\end{eqnarray} 
We are working in the collider frame, 
using natural units ($c=1$, $m_e=1$, and $\hbar=1$), 
and applying the conventional notation; $\beta\equiv v/c$, 
$\gamma\equiv1/\sqrt{1-\beta^2}$. 
$\alpha$ is the fine-structure constant, 
$\check{\alpha}$ and $\check{\gamma}^{\mu}$ are Dirac matrices 
in the Dirac representation, as in Ref.~\cite{IZ}; 
$\check{\sigma}$ are the Pauli matrices; 
and ${\rm I}_2$, $0_2$, ${\rm I}_4$, and $0_4$ are the 
2-dimensional and 4-dimensional unit and zero matrices.

We would like to consider the ultra-relativistic limit in which
\begin{eqnarray}
\beta\rightarrow 1, \ \ \ \gamma\gg b, r_{\perp}\label{url}.
\end{eqnarray}
Eq.~(\ref{dirac}) does not simplify in this limit 
in a straight-forward way because, 
for any given time, the 
long-range behavior of the interaction terms $\hat{H}_A$ and 
$\hat{H}_B$ is independent of 
$\gamma$ as $z\rightarrow\pm\infty$ (see Fig.\ \ref{lorentz_int}). 
A simple ultra-relativistic limit can be obtained by first 
applying a gauge transformation in order to 
remove this long-range tail of 
the interaction.
The gauge-transformed wave function
$|\Psi(\vec{r},t)\rangle$
is defined by
   \begin{eqnarray}
|\Psi(\vec{r},t)\rangle \equiv U(z,t) |\Phi(\vec{r},t)\rangle,
\end{eqnarray}
   \begin{eqnarray}
U(z,t) \equiv \exp && \left\{ 
i Z_A \alpha \ln \left[ -\gamma (t-z) +\sqrt{b^2 
+\gamma^2(t-z)^2}\right] \right. 
\nonumber \\
+&&\left. i Z_B \alpha \ln \left[ +\gamma (t+z) +\sqrt{b^2 
+\gamma^2(t+z)^2}\right] \right\}.
\label{newgauge}
\end{eqnarray}
The Dirac equation for $|\Psi(\vec{r},t)\rangle$ is obtained 
from Eq.~(\ref{dirac}),
   \begin{equation}
i\frac{\partial}{\partial t} |\Psi(\vec{r},t)\rangle =
\left[ \hat{H}_0 + \hat{W}_A(t) + \hat{W}_B(t) \right] 
|\Psi(\vec{r},t)\rangle \label{gt_dirac}
\end{equation}
where the new interaction terms are
   \begin{eqnarray}
\hat{W}_A(t)&=& ({\rm I}_4-\beta\check{\alpha}_z)
\frac{-Z_A\alpha}
{\sqrt{(\vec{r}_{\perp}-\vec{b})^2/\gamma^2+(z-\beta t)^2}} 
\nonumber \\
&&-({\rm I}_4-\check{\alpha}_z) \frac{-Z_A\alpha}
{\sqrt{\vec{b}^2/\gamma^2+(z- t)^2}},
\label{fullW_A}
\\
\hat{W}_B(t)&=& ({\rm I}_4+\beta\check{\alpha}_z)
\frac{-Z_B\alpha}
{\sqrt{(\vec{r}_{\perp}+\vec{b})^2/\gamma^2+(z+\beta t)^2}}
\nonumber \\
&&-({\rm I}_4+\check{\alpha}_z) \frac{-Z_B\alpha}
{\sqrt{\vec{b}^2/\gamma^2+(z + t)^2}}.  
\label{fullW_B}
\end{eqnarray} 
Figure \ref{eichler_int} demonstrates the short-range character of 
this gauge-transformed interaction.
Similar gauges have been used in Refs.~\cite{TE90,BRW} which
reduce in the limit $\beta\rightarrow 1$ to Eq.~(\ref{newgauge}).  
Unlike Eq.~(\ref{dirac}),
the gauge-transformed equation, Eq.~(\ref{gt_dirac}), 
has a simple ultra-relativistic 
limit\cite{Baltz97,Ja92}. 
In the limits of Eq.~(\ref{url})
(large $\gamma$, small $r_{\perp}$, and small impact parameter $b$,),
$\hat{W}_A$ ( $\hat{W}_B$) has a sharp, delta-function dependence
on $ t-z$ ($ t+z$)\cite{Baltz97,Ja92,RR84}, 
(see appendix \ref{delta}), i.e.\ 
\begin{eqnarray}
\hat{W}_A\rightarrow ({\rm I}_4-\check{\alpha}_z) 
Z_A \alpha \delta (t-z) \ln \left[ 
\frac{(\vec{r}_{\perp}-\vec{b})^2}{b^2} \right] 
\label{W_A}, \\
\hat{W}_B\rightarrow ({\rm I}_4+\check{\alpha}_z) 
Z_B \alpha \delta (t+z) \ln \left[ 
\frac{(\vec{r}_{\perp}+\vec{b})^2}{b^2} \right]
\label{W_B}.
\end{eqnarray}

Consider the physical nature of this limit.
A $\delta$ function over time alone would
indicate a sudden interaction of the ions with the 
vacuum. 
In the gauge-transformed equation, Eq.~(\ref{gt_dirac}), with the 
interactions of Eqs.~(\ref{W_A}) and (\ref{W_B}), as they move, 
the ions are continuously interacting with the vacuum. 
Naturally,
this interaction is singular on the trajectories of the ions, 
as it was before the ultra-relativistic limit has been taken;
but an additional singularity is induced
in the ultra-relativistic limit
by the extreme Lorentz contraction of the field.
In this limit, 
the interaction is infinite on the two planes perpendicular 
to the ions trajectories, and vanishes elsewhere.
In the following, we calculate pair production amplitudes
using Eq.\ (\ref{gt_dirac}) with the interactions in Eqs.\
(\ref{W_A}) and (\ref{W_B}).
The region of large $r_\perp $ is not properly accounted
for in this treatment,
but contributions from this region to pair production 
are assumed to be small.

The interactions in Eqs.\ (\ref{W_A}) and (\ref{W_B}) have zero 
range in the longitudinal direction and a logarithmic behavior in the 
transverse direction, similar to the potential of a line of 
charge. 
In the limit $\beta\rightarrow 1$, the two ions 
are moving at the speed of light and thus
the interaction planes described above
coincide with the light fronts, given 
by $z=\pm t$ (see Fig.\ \ref{lightfronts_def}).
Finally, we note that $({\rm 
I}_4\pm\check{\alpha}_z)/2$ are orthonormal projection 
operators. The 4-Dirac spinor wave function of the 
electron can be decomposed into two orthogonal components,
   \begin{eqnarray}
&& |\Psi_+(\vec{r},t)\rangle \equiv \frac{1}{2}({\rm I}_4 + 
\check{\alpha}_z) |\Psi(\vec{r},t)\rangle \\
&& |\Psi_-(\vec{r},t)\rangle \equiv \frac{1}{2}({\rm I}_4 -
\check{\alpha}_z) |\Psi(\vec{r},t)\rangle.
\end{eqnarray}
Each ion interacts directly only with one of 
these components; 
$Z_A$ with $|\Psi_-(\vec{r},t)\rangle$ and $Z_B$ with 
$|\Psi_+(\vec{r},t)\rangle$.

\section{Asymptotic solutions and transition amplitudes}

In scattering theory, characterized by free initial and
final states, a complete solution is generally given
by the set of asymptotic transition amplitudes
between plane waves, (the S-matrix).
In this section, we define the transition amplitudes, $S^{(j)}_k$,
for the electronic spinor wave function $ | \Phi \rangle$,
i.e.\ for Eq.\ (\ref{dirac}),
and the transition amplitudes, $A^{(j)}_k$, for the
gauge transformed wave function $ | \Psi \rangle$, i.e.\
for Eq.\ (\ref{gt_dirac}), using, as usual,
initial and final plane wave states.
Use of an initial condition of a single plane wave for 
Eq.\ (\ref{dirac}) is somewhat subtle since the theory
is never free due to long-range Coulomb effects\cite{eichler}.
This is an important point when
integrating the amplitudes to obtain predictions for
physical observables.
One should properly account for actual initial and final states 
for a given experiment.

A complete set of solutions of the free Dirac equation
is given by the Dirac plane waves; $\{|\chi_p(\vec{r},t)\rangle\}$.
Each plane wave is characterized by three continuous and two 
discrete quantum numbers; namely, 
the three components of the momentum, $\vec{p}$, 
the sign of the energy, (we use $\lambda_p=0$ for positive energy 
electrons, $\lambda_p=1$ for negative energy electrons,) 
and the spin, ($|s_p\rangle=|+\rangle$ for spin up and 
$|s_p\rangle=|-\rangle$ for spin down,) 
$p\equiv\{\vec{p},\lambda_p,s_p\}$.
The plane waves satisfy,
   \begin{eqnarray}
&&\hat{H}_0 |\tilde{\chi}_p(\vec{r})\rangle = E_p 
|\tilde{\chi}_p(\vec{r})\rangle, \\
&&|\chi_p(\vec{r},t) \rangle \equiv 
\exp (-i E_p t) |{\tilde{\chi}_p(\vec{r})}\rangle, \\
&&E_p= (-1)^{\lambda_p} \varepsilon_p, \ \ \varepsilon_p \equiv 
\sqrt{p^2+1}\label{shell1},
\end{eqnarray}
where Eq.~(\ref{shell1}) is the condition for being on the 
energy shell. An explicit form is given, for 
example, by\cite{eichler}, and in our notation by
   \begin{eqnarray}
|\tilde{\chi}_p(\vec{r})\rangle &=& 
\frac{(2\pi)^{-3/2}}{\sqrt{2\varepsilon_p(1+\varepsilon_p)}}
\exp(i \vec{r}\cdot\vec{p}) 
\nonumber \\
&&\times
\left(
\begin {array}{c}
0_2 \ \ -{\rm I}_2 \\
{\rm I}_2 \ \ \ \ 0_2
\end{array}
\right)^{\lambda_p}
\left(
\begin {array}{c}
(1+\varepsilon_p) \ |s_p\rangle \\
\vec{\check{\sigma}}\cdot\vec{p} \ \ |s_p\rangle
\end{array}
\right)
\label{planewave}  \\
&\equiv& 
\exp(i \vec{r}\cdot\vec{p}) |u_p \rangle \; .
\end{eqnarray}

We define
the solution $|\phi^{(j)}(\vec{r},t)\rangle$ of 
Eq.~(\ref{dirac}) by the initial condition,
   \begin{eqnarray}
\lim_{t_i\rightarrow -\infty}|\phi^{(j)}(\vec{r},t_i)\rangle
=|\chi_j(\vec{r},t_i)\rangle\label{initialphi}.
\end{eqnarray}
The asymptotic transition amplitude $S^{(j)}_k$ is then   
given by
   \begin{eqnarray}
S^{(j)}_k \equiv \lim_{t_f\rightarrow \infty}
\langle \chi_k(\vec{r},t_f) | \phi^{(j)}(\vec{r},t_f)\rangle,
\label{Samplitude}
\end{eqnarray}
where the bra-ket stands, as usual in non-relativistic and 
non-covariant notation, for integration over all space 
$\vec{r}$ at a given time.
Likewise, $|\psi^{(j)}(\vec{r},t)\rangle$ is defined as the 
solution of Eq.~(\ref{gt_dirac}) with the initial 
condition,
   \begin{eqnarray}
\lim_{t_i\rightarrow -\infty}|\psi^{(j)}(\vec{r},t_i)\rangle
=|\chi_j(\vec{r},t_i)\rangle\label{initialpsi},
\end{eqnarray}
and the asymptotic transition amplitude 
$A^{(j)}_k$ is given by
   \begin{eqnarray}
A^{(j)}_k \equiv \lim_{t_f\rightarrow \infty}
\langle \chi_k(\vec{r},t_f) | \psi^{(j)}(\vec{r},t_f)\rangle.
\label{amplitude}
\end{eqnarray}

The initial condition for Eq.\ (\ref{dirac}), Eq.~(\ref{initialphi}),
and the initial condition for Eq.\ (\ref{gt_dirac}), Eq.~(\ref{initialpsi}),
correspond to different initial physical states, as they are
not related by the gauge transformation in Eq.\ (\ref{newgauge}).
A similiar comment is true for the final states used
in defining the amplitudes in Eqs.\ (\ref{Samplitude}) and (\ref{amplitude}).
In general, $S^{(j)}_k$ and $A^{(j)}_k$ 
are completely different amplitudes. 
They are related to each 
other by the gauge transformation of Eq.~(\ref{newgauge}) 
in the following way,
   \begin{eqnarray}
S^{(j)}_k \equiv \sum_{p} \sum_{q}
&&\langle \chi_k(\vec{r},t_f) | U^{\dagger}(z,t_f) |
\chi_p(\vec{r},t_f)\rangle \nonumber \\
\times A^{(q)}_p \  
&&\langle \chi_q(\vec{r},t_i) | U(z,t_i) |
\chi_j(\vec{r},t_i)\rangle \label{AS},
\end{eqnarray}
where $\sum_{p}$ stands for integration and summation over 
all the quantum numbers, $p=\{\vec{p},\lambda_p,s_p\}$. 
This relation is based on the completeness of 
the plane-waves basis set and 
should be 
questioned 
if applied with a truncated basis calculation\cite{Ru90,Ru93,EB96}.

A relation like Eq.~(\ref{AS}) holds between any 
two amplitudes which are related by a
gauge transformation. 
If the gauge transformation $U$ had been otherwise defined
so that it would become unity for asymptotic times, 
(i.e.\  $ U \rightarrow 1 $ for $ t_i\rightarrow -\infty $
and $ t_f\rightarrow \infty $),
the orthonormality of the plane-waves
would eliminate the double sum in Eq.\ (\ref{AS}), 
and the {\it asymptotic} transition amplitudes 
$S^{(j)}_k $ and $A^{(j)}_k $ would be identical.
Gauge transformations which share this property have
been described as exhibiting 
{\it asymptotic gauge invariance}\cite{Ru93}.
Likewise, if the gauge transformation $U$ would be 
independent of $\vec{r}$ for asymptotic times, 
(i.e.\  $U \rightarrow \exp(iC_i)$ for $ t_i\rightarrow -\infty$, and
$ U \rightarrow \exp(iC_f) $ for $ t_f\rightarrow +\infty$,
where $C_i$ and $C_f$ are real numbers),
then the double sum in Eq.\ (\ref{AS}) would again be
eliminated and the matrix elements of $U$ in Eq.\ (\ref{AS})
would contribute only a single constant phase at asymptotic times.
In this case, the asymptotic transition probabilities derived from
$S^{(j)}_k $ and $A^{(j)}_k $ would be identical \cite{BRW}.

The specific gauge transformation used here, Eq.~(\ref{newgauge}), 
is not independent of space at asymptotic times,
as is shown in appendix \ref{z/t}, and, as a result,
does not exhibit asymptotic gauge invariance for
the amplitudes or the probabilities derived from them.
However, this gauge does have
an additional 
property which relates $S^{(j)}_k$ and $A^{(j)}_k$ in a 
way more useful than Eq.\ (\ref{AS}). 
It is shown in appendix \ref{z/t} that
in the special case of symmetric collisions, with 
$Z_A=Z_B$, 
$S^{(j)}_k$ can be expressed as a series expansion 
in powers of $1/t_f$ and $1/|t_i|$
whose zero-order term is $A^{(j)}_k$.

\section{The sharp Dirac equation in the light-fronts 
representation}

In this section, the {\it sharp} Dirac equation, 
Eq.~(\ref{gt_dirac}), with the limiting
form of the interaction in Eqs.\ (\ref{W_A}), and (\ref{W_B}),
will be further simplified by changing into light-front variables and 
by introducing a new representation for the Dirac spinors, the 
{\it light-fronts representation}. 
This is an appropriate choice of variables and representation,
since, in the ultra-relativistic limit of Eq.\ (\ref{url}),
the interactions are confined to the light fronts.

\subsection{Definitions and notations}
In terms of light-front 
variables, space-time and energy-momentum are 
described by the 4-vectors $(\vec{r}_{\perp}, \tau_+, 
\tau_-)$ and $(\vec{p}_{\perp}, p_+, p_-)$, where
   \begin{eqnarray}
&&\tau_{\pm}\equiv (t\pm z)/2 \\
&&p_{\pm}\equiv E_p\pm p_z \\
&& p_+ p_- = 1 + {p}_{\perp}^2 \label{shell2}
\end{eqnarray}
The sign and absolute value of $(p_++p_-)/2$
are $\lambda_p$ and $\varepsilon_p$, respectively.
Equation (\ref{shell2}) like Eq.~(\ref{shell1}) defines the 
energy-shell.
These variables were often used previously for quantization on one of 
the two light fronts, $\tau_+=0$ or $\tau_-=0$\cite{BK71}. 
For the problem considered here, it is useful to keep the  
symmetry between $\tau_+$ and $\tau_-$. 

The projection operators 
$({\rm I}_4\pm\check{\alpha}_z)/2$ 
acquire a simple form and the interaction is 
diagonalized by
introducing the {\it light-fronts 
representation} for the Dirac matrices, 
   \begin{eqnarray}
&&\gamma^{\mu}_{\rm light-fronts}=
\Lambda \gamma^{\mu}_{\rm Dirac} 
\Lambda^{\dagger}, \\
&& \Lambda \equiv \frac{1}{\sqrt{2}}\left(
\begin {array}{c} 
{\rm I}_2 \ \ \ \ \hat{\sigma}_z \\
{\rm I}_2 \ - \hat{\sigma}_z
\end{array} 
\right)\label{lf}, \\
&&\Lambda \check{\alpha}_z
\Lambda^{\dagger} = \left(
\begin {array}{c}
{\rm I}_2 \ \ \ \ 0_2 \\
0_2 \ -{\rm I}_2 
\end{array}
\right),
\\
&&\Lambda 
\left[ \frac{1}{2}({\rm I}_4+\check{\alpha}_z)\right] 
\Lambda^{\dagger} = \left(
\begin {array}{c}
{\rm I}_2 \ 0_2 \\
0_2 \ 0_2
\end{array}
\right),
\\
&&\Lambda 
\left[ \frac{1}{2}({\rm I}_4-\check{\alpha}_z)\right]
\Lambda^{\dagger} = \left(
\begin {array}{c}
0_2 \ 0_2 \\
0_2 \ {\rm I}_2
\end{array}
\right),
\\
&&\Lambda \check{\vec{\alpha}}_{\perp}
\Lambda^{\dagger} = \left(
\begin {array}{c}
0_2 \ - \check{\vec{\omega}} \\
\check{\vec{\omega}} \ \ \ \ \ 0_2
\end{array}
\right),
\\
&& \check{\vec{\omega}}\equiv 
(-\check{\sigma}_y , \check{\sigma}_x).
\end{eqnarray}

With this notation, the gauge-transformed two-center Dirac 
equation in 
the sharp ultra-relativistic limit in the light-fronts  
representation is
   \begin{eqnarray}
&&\left(
\begin {array}{c}
i \partial_{\tau_+} |G_+\rangle \\ 
i \partial_{\tau_-} |G_-\rangle
\end{array}
\right) =
\left(
\begin {array}{c}
\delta (\tau_+)B(\vec{r}_{\perp}) 
\ \ \ \ \ \ \ \ \ \hat{h}_0 \\
\hat{h}_0^{\dagger} \ \ \ \ \ \ \ \ 
\delta (\tau_-) A(\vec{r}_{\perp}) 
\end{array}
\right) \left(
\begin {array}{c}
 |G_+\rangle \\
 |G_-\rangle
\end{array}
\right),
\nonumber \\ 
&&\label{sharp}
 \end{eqnarray}
where $|G_+\rangle$ and $|G_-\rangle$ are
the upper and lower bi-spinor components
of the Dirac wave function in the light-fronts 
representation
   \begin{eqnarray}
\left(
\begin {array}{c}
|G_+\rangle \\ |G_-\rangle
\end{array}
\right) \equiv \Lambda |\Psi\rangle,
\end{eqnarray}
and 
   \begin{eqnarray}
&& \hat{h}_0 \equiv {\rm I}_2 - i 
\check{\vec{\omega}}\cdot 
\hat{\vec{p}}_{\perp}\label{hatH},\\
&& A(\vec{r}_{\perp},\vec{b})\equiv Z_A \alpha \ln \left[ 
\frac{(\vec{r}_{\perp}-\vec{b})^2}{b^2} \right]\label{A}, \\
&& B(\vec{r}_{\perp},\vec{b})\equiv Z_B \alpha \ln \left[ 
\frac{(\vec{r}_{\perp}+\vec{b})^2}{b^2} \right]\label{B}. 
\end{eqnarray}
The upper and lower 
bi-spinors are coupled by the free Hamiltonian. Each interacts 
directly with the external field of one ion and 
feels the field of the other ion 
through its coupling to the other bi-spinor.

Equation (\ref{sharp}) has no discontinuities in the transverse 
direction. It is therefore useful to Fourier transform its 
solution with respect to $\vec{r}_{\perp}$.
Two mixed bi-spinors wave-functions, 
$|g_{\pm}(\vec{q}_{\perp};\tau_+,\tau_-)\rangle$,
are then defined by
   \begin{eqnarray}
&&|G_{\pm}(\vec{r}_{\perp},\tau_+,\tau_-)\rangle
\equiv \int d\vec{q}_{\perp} 
e^{i\vec{r}_{\perp}\cdot\vec{q}_{\perp}}
|g_{\pm}(\vec{q}_{\perp};\tau_+,\tau_-)\rangle.
\label{mixed}
\end{eqnarray}
$|g_+\rangle$ and $|g_-\rangle$, like $|G_+\rangle$ and 
$|G_-\rangle$, are coupled by the free Hamiltonian. 

\subsection{Free Dirac equation off the light fronts}
Off the light fronts, i.e.\ for $\tau_+\neq 0$ and $\tau_-\neq 0$, 
the wave function satisfies the free Dirac equation and
Eq.~(\ref{sharp}) reduces to two coupled equations for 
the mixed bi-spinors 
$|g_{\pm}(\vec{q}_{\perp};\tau_+,\tau_-)\rangle$.
   \begin{eqnarray}
&&i \frac{\partial}{\partial\tau_+}
|g_+\rangle =
({\rm I}_2 - i \check{\vec{\omega}}\cdot\vec{q}_{\perp})
|g_-\rangle \label{g+=},\\
&&i \frac{\partial}{\partial\tau_-}
|g_-\rangle =
({\rm I}_2 + i \check{\vec{\omega}}\cdot\vec{q}_{\perp})
|g_+\rangle \label{g-=}.
\end{eqnarray}
As usual, the second-order equations decouple
   \begin{eqnarray}
\frac{\partial^2}{\partial\tau_+\partial\tau_-}
|g_{\pm}\rangle 
= - (1+{q}^2_{\perp})
|g_{\pm}\rangle, 
\end{eqnarray}
where use was made of
   \begin{eqnarray}
({\rm I}_2 - i \check{\vec{\omega}}\cdot\vec{q}_{\perp})
({\rm I}_2 + i \check{\vec{\omega}}\cdot\vec{q}_{\perp})
= (1+{q}^2_{\perp}) {\rm I}_2 .
\end{eqnarray}

A solution to Eqs.~(\ref{g+=}, \ref{g-=}) is given, for 
example, by the plane waves of Eq.~(\ref{planewave}) 
which in the light-fronts representation 
are given by
   \begin{eqnarray}
&&\left(
\begin {array}{c}
|F^{p}_+\rangle \\ |F^{p}_-\rangle
\end{array}
\right) \equiv \Lambda |\chi_p (\vec{r},t)\rangle, 
\label{mpwb}\\
&&|F^{p}_{\pm}\rangle
\equiv \int d\vec{q}_{\perp} 
e^{i\vec{r}_{\perp}\cdot\vec{q}_{\perp}}
|f^{p}_{\pm}(\vec{q}_{\perp};\tau_+,\tau_-)\rangle,
\label{mpw} \\
&&|f^{p}_{\pm}(\vec{q}_{\perp};\tau_+,\tau_-)\rangle
= \delta (\vec{q}_{\perp}-\vec{p}_{\perp})
e^{-i (\tau_-p_+ + \tau_+p_-)}
|\Gamma^{p}_{\pm}\rangle.
\label{mpwa}
\end{eqnarray}
The bi-spinors, $|\Gamma^{p}_{\pm}\rangle$, 
(the upper and lower parts of $\Lambda |u_p\rangle$),
   \begin{eqnarray}
|\Gamma^{p}_{\pm}\rangle &=&
\frac{(2\pi)^{-3/2}}{2\sqrt{\varepsilon_p(1+\varepsilon_p)}}
\times \nonumber \\ &&
\left[{\rm I}_2 \left(1 + (-1)^{\lambda_p}p_{\pm}\right)
\pm i \check{\vec{\omega}}\cdot\vec{p}_{\perp}\right]
(\pm \check{\sigma}_z)^{\lambda_p}|s_p\rangle ,
\label{Gamma}
\end{eqnarray}
satisfy the simple relation
   \begin{eqnarray}
|\Gamma^{p}_{-}\rangle =
\frac{
{\rm I}_2 - i \check{\vec{\omega}}\cdot\vec{p}_{\perp}
}{p_+}
|\Gamma^{p}_{+}\rangle. 
\end{eqnarray}
These plane waves solve Eq.~(\ref{sharp}) off the light fronts in the 
limits $t\rightarrow\pm\infty$. 
They do not solve it for finite $t$, when $\vec{p}_\perp$ 
is no longer a good quantum number,
as the singular interaction with the ions makes the wave 
function discontinuous at the light fronts.

\subsection{The discontinuity across the light fronts}

It is standard procedure in wave-mechanics to form piece-wise solutions 
by satisfying continuity relations at the boundaries between free regions. 
It was shown in Refs.~\cite{Baltz97,Ja92} that a $\delta$-function 
singular interaction at a light front results in  a
discontinuity in the electron wave function which is given by a 
space-dependent phase shift. The proof is reviewed in appendix 
\ref{discont}, where it is shown that for our case of Eq.~(\ref{sharp}),
the discontinuity is 
   \begin{eqnarray}
&& |G_+(\tau_+=0^+)\rangle =
e^{-i B(\vec{r}_{\perp},\vec{b})}
|G_+(\tau_+=0^-)\rangle,
\label{dGB} \\
&& |G_-(\tau_-=0^+)\rangle =
e^{-i A(\vec{r}_{\perp},\vec{b})}
|G_-(\tau_-=0^-)\rangle.
\label{dGA}
\end{eqnarray}
These phase shifts are derived from Eq.~(\ref{sharp}) in general 
for any $A(\vec{r}_{\perp},\vec{b})$ and $B(\vec{r}_{\perp},\vec{b})$, 
i.e.\ any functional dependence on the perpendicular coordinate.
Here, $A(\vec{r}_{\perp}, \vec{b})$ and $B(\vec{r}_{\perp}, \vec{b})$ 
are given by Eqs.~(\ref{A}, \ref{B}). 

Due to this space dependent phase-shift, the transverse 
momentum is not conserved 
and the Fourier components of Eq.~(\ref{mixed}) are mixed
when the singularities at the light fronts are crossed,
   \begin{eqnarray}
|g_+(\vec{q}_{\perp};\tau_+=0^+)\rangle &=&
\int d \vec{p}_{\perp}
Q_{Z_B} (\vec{p}_{\perp}-\vec{q}_{\perp},-\vec{b})
\nonumber \\ &&
\times
|g_+(\vec{p}_{\perp};\tau_+=0^-)\rangle, \label{dfb}\\
|g_-(\vec{q}_{\perp};\tau_-=0^+)\rangle &=&
\int d \vec{p}_{\perp}
Q_{Z_A} (\vec{p}_{\perp}-\vec{q}_{\perp}, \vec{b})
\nonumber \\ &&
\times
|g_-(\vec{p}_{\perp};\tau_-=0^-)\rangle, \label{dfa}
\end{eqnarray} 
where
   \begin{eqnarray}
&&Q_{Z}(\vec{\kappa},\vec{b})\equiv
\frac{1}{(2\pi)^2}\int d \vec{r}_{\perp}
\ e^{i \vec{r}_{\perp}\cdot\vec{\kappa}}
\left[\frac{(\vec{r}_{\perp}-\vec{b})^2}{b^2}\right]^{- i \alpha Z}
\label{F}.
\end{eqnarray}
Note that here $\vec{\kappa}$ and $\vec{b}$ are 
two-dimensional vectors in the $(x,y)$ plane. 
The continuity is recovered  in the limit $Z\rightarrow 0$, 
as $Q_{Z} (\vec{\kappa},\vec{b})\rightarrow \delta 
({\vec \kappa})$.
The distribution $ Q_{Z}(\vec{\kappa},\vec{b}) $ in general diverges.
This divergence is an artifact of applying the sharp limit
for the gauge-transformed interaction, Eqs.\ (\ref{W_A}) and (\ref{W_B}),
for large $ r_\perp $, i.e.\
outside its range of validity.
The properties of the distribution $Q_{Z} (\vec{\kappa},\vec{b})$ for 
finite charge are considered in appendix \ref{Q_Z}.

\section{A piece-wise solution to the sharp Dirac equation}

In this section, the formalism that was introduced in 
section IV is used to obtain the transition amplitudes
between asymptotic plane waves, $A^{(j)}_k$, defined in section III.

The singular interaction on the planes perpendicular to the 
trajectories of the ions,
cut space-time along the light fronts into four regions, as is shown in
Fig.\ \ref{light_fronts}.
A piece-wise solution is defined off the light fronts by
$|g_{\pm}
(\vec{q}_{\perp};\tau_+,\tau_-)\rangle = |g^{\rm 
(i)}_{\pm}(\vec{q}_{\perp};\tau_+,\tau_-)\rangle$,
where 
(i)$=$ I for $\tau_+<0$ and $\tau_-<0$, 
(i)$=$ II for $\tau_+>0$ and $\tau_-<0$, 
(i)$=$ III for $\tau_+<0$ and $\tau_->0$, and
(i)$=$ IV for $\tau_+>0$ and $\tau_->0$.
In each region, the wave function is continuous and 
solves the local free Dirac equation.
At any time, except for $t\rightarrow\pm\infty$, 
the wave function extends in space 
through three (or two, at $t=0$) of these regions. 
The solution presented here is not complete in the sense that it
does not include the solution on the light fronts;
$ \tau_+ =0$ and $ \tau_- =0 $ are excluded.
The physics on the light fronts may contribute to bound-free pair
production.
Thus, our present work is limited to free pair production.

\subsection{Initial condition and intermediate states}

Consider the initial condition, Eq.~(\ref{initialpsi}), 
of a single plane wave with the quantum numbers 
$j=\{\vec{j},\lambda_j,s_j\}$,
or, using light-front variables, 
$j=\{\vec{j}_{\perp},j_+,j_-,s_j\}$,
with the constraint $j_+j_-=1+{j}_{\perp}^2$.
The continuity off the light 
fronts gives the solution in region I,
   \begin{eqnarray}
&&|g^{\rm I}_{\pm}(\vec{q}_{\perp})\rangle
= \delta (\vec{j}_{\perp}-\vec{q}_{\perp})
e^{-i (\tau_-j_+ + \tau_+j_-)}
|\Gamma^{j}_{\pm}\rangle,
\end{eqnarray}
where the bi-spinors $|\Gamma^{j}_{\pm}\rangle$ are 
defined as in Eq.~(\ref{Gamma}).

The solution in regions II and III is obtained
by first applying Eq.~(\ref{dfb}) for the discontinuity 
across $\tau_+=0$ and Eq.~(\ref{dfa}) for the 
discontinuity across $\tau_-=0$ and then solving the coupled 
equations (\ref{g+=}, \ref{g-=}) inside each of the 
intermediate space-time regions. We obtain in region II
   \begin{eqnarray}
|g^{\rm II}_{+}(\vec{q}_{\perp})\rangle
&=& 
\exp \left[-i\tau_-j_+
-i\tau_+\left(\frac{1+q^2_{\perp}}{j_+}\right)\right]
 \nonumber \\ && \times 
Q_{Z_B} (\vec{j}_{\perp}-\vec{q}_{\perp},-\vec{b})
\ |\Gamma^{j}_{+}\rangle, 
 \nonumber \\
|g^{\rm II}_{-}(\vec{q}_{\perp})\rangle
&=& \left(\frac {I_2+i 
\check{\vec{\omega}}\cdot\vec{q}_{\perp}}{j_+}\right)
\ |g^{\rm II}_{+}(\vec{q}_{\perp})\rangle \label{II},
\end{eqnarray}
and in region III,
   \begin{eqnarray}
|g^{\rm III}_{-}(\vec{p}_{\perp})\rangle
&=& 
\exp \left[-i\tau_+j_-
-i\tau_-\left(\frac{1+p^2_{\perp}}{j_-}\right)\right]
\nonumber \\ && \times 
Q_{Z_A} (\vec{j}_{\perp}-\vec{p}_{\perp},\vec{b})
\ |\Gamma^{j}_{-}\rangle, 
 \nonumber \\
|g^{\rm III}_{+}(\vec{p}_{\perp})\rangle
&=& \left(\frac {I_2-i 
\check{\vec{\omega}}\cdot\vec{p}_{\perp}}{j_-}\right)
\ |g^{\rm III}_{-}(\vec{q}_{\perp})\rangle \label{III} .
\end{eqnarray}

It is now apparent why the Fourier transform with respect to 
$\vec{r}_{\perp}$ and the definition of 
$|g_{\pm}(\vec{q}_{\perp};\tau_+,\tau_-)\rangle$
in Eq.~(\ref{mixed}) were needed. 
The simple discontinuity condition (\ref{dGB}) at $\tau_+=0$
applies only to $|G_+\rangle$. The other bi-spinor $|G_-\rangle$ 
is influenced indirectly by the field at $\tau_+=0$
through its coupling to $|G_+\rangle$. 
Likewise, at $\tau_-=0$ the simple discontinuity condition (\ref{dGA}) for 
$|G_-\rangle$ induces a non-trivial change in $|G_+\rangle$.
The coupling between 
$|G_+\rangle$ and $|G_-\rangle$ in free space on either sides of the 
singular interaction is best described by Eqs.~(\ref{g+=},\ref{g-=}) for 
their Fourier components with respect to
$\vec{r}_{\perp}$.
Thus, while the discontinuity conditions (\ref{dfb},\ref{dfa}) for 
$|g_\pm\rangle$  
seem more complicated than the discontinuity conditions 
(\ref{dGB},\ref{dGA})
for $|G\pm \rangle$, using $|g_\pm\rangle$ allows for a simple 
derivation of the complete spinor wave function in regions II and III.

It is a well known fact that
two ions are needed in order to create an electron-positron pair. 
This can also be seen here. 
In the presence of ion B alone, for example, Eqs.~(\ref{II}) give 
the solution for $\tau_+>0$, including the asymptotic solution at 
$t_f\rightarrow\infty$. Projection on a plane wave then gives a 
conservation law for the positive light-front momentum. Likewise, in the 
presence of ion A alone the negative light front momentum is conserved.
   \begin{eqnarray}
A_k^{(j)}(Z_A=0) \propto \delta (k_+-j_+), \\
A_k^{(j)}(Z_B=0) \propto \delta (k_--j_-).
\end{eqnarray}
A direct result from $k_+=j_+$ or $k_-=j_-$ is that the sign of the 
energy of 
the electron is the same before and after the collision. 
Thus, our formalism satisfies the known result that the passage
of a uniformly moving charge does not induce a transition changing
the sign of the energy.
The presence of both ions defines a new region of space-time, region IV 
($\tau_\pm >0$) which is the space between the ions ($-t<z<t$) after the 
collision ($t>0$), i.e.\ when the ions are already moving apart. It is 
shown below that the sign of the energy can
change and pairs may be created in
transitions from the initial state in region I ($t_i\rightarrow - \infty$)
to the final state in region IV ($t_f\rightarrow \infty$).

The solution of the free Dirac equation in region IV is complicated by 
the non-trivial boundary conditions on the light fronts.
Applying Eq.~(\ref{dfb}) again for the discontinuity 
across $\tau_+$ and Eq.~(\ref{dfa}) for the 
discontinuity across $\tau_-$,
we cross from regions II and III into region IV
to obtain on the hyper-surfaces
adjacent to the light fronts,
   \begin{eqnarray}
|g^{\rm IV}_{-}(\vec{k}_{\perp}; \tau_-=0^+)\rangle
&=& \int d \vec{q}_{\perp}
\exp
\left[-i\tau_+\left(\frac{1+q^2_{\perp}}{j_+}\right)\right]
\nonumber \\ && \times
\ Q_{Z_A} (\vec{q}_{\perp}-\vec{k}_{\perp},\vec{b})
\nonumber \\ && \times
\ Q_{Z_B} (\vec{j}_{\perp}-\vec{q}_{\perp},-\vec{b})
\nonumber \\ && \times
\left(\frac {I_2+i 
\check{\vec{\omega}}\cdot\vec{q}_{\perp}}{j_+}\right)
\ |\Gamma^{j}_{+}\rangle, 
\label{bc-}
\\ 
|g^{\rm IV}_{+}(\vec{k}_{\perp}; \tau_+=0^+)\rangle
&=& \int d \vec{p}_{\perp}
\exp
\left[-i\tau_-\left(\frac{1+p^2_{\perp}}{j_-}\right)\right]
\nonumber \\ && \times
\ Q_{Z_B} (\vec{p}_{\perp}-\vec{k}_{\perp}-\vec{b})
\nonumber \\ && \times
\ Q_{Z_A} (\vec{j}_{\perp}-\vec{p}_{\perp},\vec{b})
\nonumber \\ && \times
\left(\frac {I_2-i 
\check{\vec{\omega}}\cdot\vec{p}_{\perp}}{j_-}\right)
\ |\Gamma^{j}_{-}\rangle.
\label{bc+}
\end{eqnarray}
Instead of solving now for
$|g^{\rm IV}_{\pm}\rangle$ at any $\tau_\pm >0$, the transition
amplitudes are obtained in the next subsection by defining the transition
current and by applying Gauss' theorem for this current.

\subsection{The amplitudes}
The transition amplitudes $A^{(j)}_k$
were defined in Eq.~(\ref{amplitude}),
   \begin{eqnarray}
A^{(j)}_k \equiv 
\lim_{t_f\rightarrow \infty}
\int d \vec{r}
\ \chi_k^{\dagger}(\vec{r},t_f) 
\ \psi^{(j)}(\vec{r},t_f).
\end{eqnarray}
The integrand is a component of a 4-vector current density,
which is a conserved quantity (see appendix \ref{current}). 
This {\it transition current} \cite{Ru90}, $(\vec{J}^{(k,j)}, J_0^{(k,j)})$, 
is defined by 
   \begin{eqnarray}
&&J_0^{(k,j)}\equiv\chi_k^{\dagger}\ \psi^{(j)} \nonumber \\
&&\vec{J}^{(k,j)} \equiv \ \chi_k^{\dagger} \ \check{\vec{\alpha}} 
\ \psi^{(j)}. 
\label{tcurrent}
\end{eqnarray}
An equivalent form for the transition current in terms of
light-fronts representation wave-functions includes
   \begin{eqnarray}
J_{\pm}^{(k,j)} \equiv  J_0^{(k,j)}\pm J_z^{(k,j)} 
\ =\ 2 \ F_{\pm}^{k \dagger} \ G_{\pm}^{(j)} 
\label{currents}. 
\end{eqnarray}

It is now possible to use Gauss' theorem 
on the hyper-surface of the inner border of region IV 
to show that (see appendix \ref{gauss_app})
   \begin{eqnarray}
A^{(j)}_k &\equiv& 
\lim_{t_f\rightarrow \infty}
\int d \vec{r}_{\perp} \int^{\infty}_{-\infty} d z
\ J_0^{(k,j)}(\vec{r},t_f) \nonumber \\
&=& 2 \int d \vec{r}_{\perp}
\int_{0^+}^{\infty} d \tau_-
\ J_+^{(k,j)}(\vec{r}_{\perp},\tau_+=0^+,\tau_-) \nonumber \\
&& -2 \int d \vec{r}_{\perp}
\int_{0^+}^{\infty} d \tau_+
\ J_-^{(k,j)}(\vec{r}_{\perp},\tau_+,\tau_-=0^+)\label{gauss} .
\end{eqnarray}
The transition currents $J_{\pm}^{(k,j)}$
are calculated from the results of 
the last subsection
by using Eqs.~(\ref{mixed},\ref{mpwb}-\ref{mpwa},\ref{currents}).
     \begin{eqnarray}
J_{\pm}^{(k,j)} (\vec{r}_{\perp},\tau_+,\tau_-) &=& 
2 \int d \vec{p}_{\perp} \int d \vec{l}_{\perp} \exp [i 
\vec{r}_{\perp} \cdot (\vec{l}_{\perp}-\vec{p}_{\perp})] \nonumber \\
&& \times \langle f^k_{\pm}(\vec{p}_{\perp};\tau_+,\tau_-) |
g^{\rm IV}_{\pm}(\vec{l}_{\perp};\tau_+,\tau_-)\rangle .
\end{eqnarray}
Integrating over $\vec{r}_{\perp}$ and 
using the explicit expression (\ref{mpwa}) for the plane waves,
   \begin{eqnarray}
A^{(j)}_k &=& 
\int_{0^+}^{\infty} d \tau_-
e^{i \tau_- k_+} \langle \Gamma_+^k |
g^{\rm IV}_{+}(\vec{k}_{\perp};\tau_+=0^+,\tau_-)\rangle \nonumber \\
&& -2 \int_{0^+}^{\infty} d \tau_+
e^{i \tau_+ k_-} \langle \Gamma_-^k |
g^{\rm IV}_{-}(\vec{k}_{\perp};\tau_+,\tau_-=0^+)\rangle.
 \label{intgauss}
\end{eqnarray}
The amplitudes are finally obtained by substituting 
Eqs.~(\ref{bc-},\ref{bc+})
and integrating over $\tau_{\pm}$.
The integration over $\tau_{\pm}$ would have given a 
$\delta$-function conservation law for the light-front momenta, 
had it been on the complete line $-\infty<\tau_{\pm}<\infty$. Instead, 
the integrals on the half lines $0<\tau_{\pm}<\infty$ are regulated in 
the usual way with an infinitesimal small constant, $\eta$\cite{newton}.
   \begin{eqnarray}
\int_{0^+}^{\infty} d \tau \exp (i \tau \kappa) = 
\frac{- i}{\kappa +i \eta}.
\end{eqnarray}

The transition amplitudes corresponding 
to the {\it exact} 
solution of the sharp Dirac equation off the light fronts are 
   \begin{eqnarray}
A^{(j)}_k &=& \frac{(2\pi)^3}{i\pi} 
      \times \nonumber \\ &&
\left\{ \int d \vec{p}_{\perp}
\ Q_{Z_B} (\vec{p}_{\perp}-\vec{k}_{\perp},-\vec{b})
\ Q_{Z_A} (\vec{j}_{\perp}-\vec{p}_{\perp},\vec{b})
       \right. \nonumber \\ &&  \times
\frac{\langle\Gamma_+^k|
I_2-i\check{\vec{\omega}}\cdot\vec{p}_{\perp}
|\Gamma_-^j\rangle}
{j_-k_+-(1+p^2_{\perp})+i\eta(-1)^{\lambda_j}}
      \nonumber \\  &&
-\int d \vec{q}_{\perp}
\ Q_{Z_A} (\vec{q}_{\perp}-\vec{k}_{\perp},\vec{b})
\ Q_{Z_B} (\vec{j}_{\perp}-\vec{q}_{\perp},-\vec{b})
        \nonumber \\ && \left. \times 
\frac{\langle\Gamma_-^k|
I_2+i\check{\vec{\omega}}\cdot\vec{q}_{\perp}
|\Gamma_+^j\rangle}
{j_+k_--(1+q^2_{\perp})+i\eta(-1)^{\lambda_j}}
\right\}, \label{amp}
\end{eqnarray}
where the infinitesimal small, positive constant, $\eta$,
can be omitted for pair-production amplitudes
corresponding to $E_j<0$ and $E_k>0$, i.e.\ 
$j_{\pm}k_{\mp}<0$.
This is an interesting result. In the ultra-relativistic limit of 
Eq.~(\ref{url}), the asymptotic time evolution of the 
gauge-transformed electron wave function, $|\Psi(\vec{r},t)\rangle$, is 
exactly given by these amplitudes. 
An exponential, non-perturbative 
dependence on the coupling constant $\alpha Z$ appears here as non-trivial 
phases in the integral representation of the distributions $Q_{Z_A} 
(\vec{\kappa}_{\perp},\vec{b})$ and $Q_{Z_B} 
(\vec{\kappa}_{\perp},-\vec{b})$ which were defined in 
Eq.~(\ref{F}).
The two terms in Eq.~(\ref{amp}) correspond to two different 
time-orderings of the interaction with the ions. In the next subsection,
they are shown to reduce in the small-coupling perturbative limit to the 
well known two photon exchange diagrams as depicted in Fig.\ \ref{figure7}.
For vanishing charges, as could be expected,
  \begin{eqnarray}
A_k^{(j)}(Z_A=0, Z_B=0)=\delta(\vec{k}-\vec{j})\delta_{\lambda_k,\lambda_j}
\delta_{s_k,s_j}.
\end{eqnarray}

\subsection{The perturbative limit}

The small-charge perturbative-limit of the pair-production amplitude was
calculated
in Ref.~\cite{BS89a}. To leading order in $\alpha Z$ (second order),
the amplitude is given by a sum over two diagrams, where each diagram
describes a two-photon exchange process.
The second-order perturbation-theory result,
${\cal{S}}_k^{2 \ (j)}$,
for the transition amplitude between an initial
negative-energy state $j=\{\vec{j},\lambda_j=1,s_j\}$ and a final
positive-energy state $k=\{\vec{k},\lambda_k=0,s_k\}$, is given by
Eqs.~(24--32) of Ref.~\cite{BS89a}.
In the ultra-relativistic limit, $\beta\rightarrow 1$ and
large $\gamma$, the perturbative result reduces to
   \begin{eqnarray}
{\cal{S}}^{2 \ (j)}_k &=&
\int d \vec{p}_{\perp}
\exp [- i \vec{b}\cdot
(2\vec{p}_{\perp}-\vec{j}_{\perp}-\vec{k}_{\perp})]
      \nonumber \\ && \times
\frac{ i 8 \ (\alpha Z_A)(\alpha Z_B)}
{(\vec{p}_{\perp}-\vec{k}_{\perp})^2
\ (\vec{p}_{\perp}-\vec{j}_{\perp})^2}
      \nonumber \\ && \times
\frac{\langle\Gamma_+^k|
I_2-i\check{\vec{\omega}}\cdot\vec{p}_{\perp}
|\Gamma_-^j\rangle}
{j_-k_+-(1+p^2_{\perp})}
      \nonumber \\  &&
- \int d \vec{q}_{\perp}
\exp [ i \vec{b}\cdot
(2\vec{q}_{\perp}-\vec{j}_{\perp}-\vec{k}_{\perp})]
      \nonumber \\ && \times
\frac{ i 8 \ (\alpha Z_A)(\alpha Z_B)}
{(\vec{q}_{\perp}-\vec{k}_{\perp})^2
\ (\vec{q}_{\perp}-\vec{j}_{\perp})^2}
      \nonumber \\ && \times
\frac{\langle\Gamma_-^k|
I_2 + i\check{\vec{\omega}}\cdot\vec{q}_{\perp}
|\Gamma_+^j\rangle}
{j_+k_--(1+q^2_{\perp})}. \label{amp.pert}
\end{eqnarray}
A transformation to the light-fronts representation was used here to
obtain bi-spinor bra-kets from the 4-spinor bra-kets of
Ref.~\cite{BS89a}. For example, using $\Lambda^{\dagger}\Lambda\equiv 
{\rm I}_4$,
     \begin{eqnarray}
&&\langle u_k| \ (I_4-\check{\alpha}_z) \
(\check{\alpha}\cdot
\vec{p}_{\perp}+\check{\gamma}^0) \ (I_4+\check{\alpha}_z) \
| u_j\rangle
\nonumber \\ &&
\equiv 
\langle\Gamma_+^k| I_2-i\check{\vec{\omega}}\cdot\vec{p}_{\perp}
|\Gamma_-^j\rangle.
\end{eqnarray}

It is interesting to compare the {\it perturbative} result of
Eq.~(\ref{amp.pert}) to our {\it non-perturbative} result of
Eq.~(\ref{amp}).
In the small-charge limit of $\alpha Z\rightarrow 0$,
after proper regularization, the leading-order perturbative 
limit for $Q_{Z}$ from appendix \ref{Q_Z} can be used,
\begin{eqnarray}
Q_{Z}(\vec{\kappa},\vec{b}) \rightarrow \delta(\vec{\kappa})
-\frac{i \alpha Z}{\pi}
\frac{1}{\kappa^2} \exp [i \vec{b}\cdot\vec{\kappa}].
\label{pertF}
\end{eqnarray}
Direct substitution shows that in this limit the non-perturbative result of
Eq.~(\ref{amp}) exactly reproduces the perturbative result of
Eq.~(\ref{amp.pert}).

\section{Conclusions and Outlook}

We have used a gauge transformation to obtain a useful 
ultra-relativistic limit for the two-center Dirac 
equation, which allows for an exact solution off the
planes perpendicular to the ions' trajectories, 
i.e.\ off the light fronts.
In general, the amplitudes of the gauge-transformed 
Dirac equation are related to the amplitudes of the
original equation in a non-trivial way due to long-range 
Coulomb effects. 
For symmetric collisions, and 
calculations in the collider 
frame, some of these long-range effects cancel.
The two different amplitudes are then related by a 
series expansion, and, to leading order, 
they are equal.

The amplitudes were calculated here in the 
ultra-relativistic limit, assuming $\gamma$ to be large. 
No assumption was made on the value of the charge times the 
fine-structure constant $Z \alpha$. 
When taking the limit of small $Z \alpha$,
we are able to show a complete 
agreement with the ultra-relativistic limit of the 
expression obtained from standard second-order 
perturbation theory\cite{BS89a}. 
In second-order perturbation theory, pair production 
is described as a two photon exchange process in which 
each ion exchanges one photon with a negative-energy 
electron. The negative energy electron is 
kicked off its energy shell by the first interaction 
and then kicked back to the energy shell by the second ion, 
but with a positive energy. 
The two diagrams that 
contribute to the amplitude differ in the time order of these 
photon exchanges, or `kicks'.
Our result, which provides a very similar physical picture of pair 
production as a `two-kicks' process,
is obtained in the ultra-relativistic 
limit within a rather different, and completely {\it 
non-perturbative} approach. 

In our work, the electromagnetic fields of the ions are confined to
the light fronts by the extreme Lorentz contraction and by a choice of a 
particular gauge designed to remove the long-range Coulomb effects. In 
this gauge, as the velocity of the ions approaches the velocity of 
light, each ion carries with it, perpendicular to its trajectory, a
wall of singular electromagnetic interaction. 
An initial plane wave in the space between the approaching ions acquires 
a space-dependent phase shift as it is swept by this 
singular-interaction wall. A single plane wave between the ions gives a 
distribution of local plane waves in the space behind each ion. Had 
there been only one ion,
no transition would be allowed between the negative-energy continuum
and the positive-energy continuum, i.e.\ no pairs could be
produced.  Pairs are 
produced because, as the ions move past each other, the two 
phase-shift planes collide. After the peripheral collision, as the ions 
move apart, the solution in the space between them is determined by the 
non-trivial boundary conditions at the light fronts. 
The main result of this work, 
the exact integral representation for the 
free pair creation amplitudes of Eq.~(\ref{amp}), 
is finally obtained in this framework by calculating 
the transition currents flowing from the light fronts into the space 
between the separating ions. 
The two terms correspond to the two time orderings of the 
interaction of the two phase-shift walls with the electronic 
wave function. 
In the perturbative limit of a small coupling constant, 
the effect of the singular field perpendicular to each ion reduces to a 
single photon exchange. For finite charges of the ions, the perturbative 
linear dependence of the amplitudes on each charge is replaced by 
non-perturbative, nontrivial phases in our integral representation.  

Numerical evaluation of the non-perturbative effects, 
differential cross sections, and applications to 
multiple pair production will be
considered in future work. 

\section*{Acknowledgments}

The authors are happy to acknowledge helpful
conversations with Prof.\ J.\ Macek and Prof.\ M.\ R.\ Strayer.
This work was supported by the National Science Foundation 
through a grant for the Institute for Theoretical Atomic and Molecular 
Physics at Harvard University and Smithsonian Astrophysical Observatory.

\appendix 
\section{The sharp, ultra-relativistic limit }
\label{delta}

In this appendix, we will outline the derivation of the 
$\delta$-function
limit of the 
electromagnetic interactions, $W_A$ 
and $W_B$, given in (\ref{W_A}) and (\ref{W_B}) beginning
with their definitions given in (\ref{fullW_A}) and (\ref{fullW_B}),
respectively. (The same limit has been previously obtained;
see Refs.\ \cite{Baltz97,Ja92,RR84}).

In the limit of extreme ultra-relativistic collisions,
one may neglect terms in the interaction
proportional to $ \gamma^{-2} $. It is then possible to first 
set $ \beta \rightarrow 1 $ and then use
\begin{eqnarray}
\eta(\tau; a, b)&&\equiv 
\left(\frac{b^2}{\gamma^2}+\tau^2\right)^{-1/2} -
\left(\frac{a^2}{\gamma^2}+\tau^2\right)^{-1/2}
\nonumber \\
&& \stackrel{\gamma \gg a,b}{\longrightarrow} \delta(\tau) \ln \left( 
\frac{a^2}{b^2}\right).
\label {deltalimit}
\end{eqnarray}
It is easy to verify that in the limit $\gamma \gg a,b$,
\begin{eqnarray}
\int_{-\infty}^{\infty} \ d \tau \  
\eta(\tau; a, b) {\rightarrow} \ln \left(
\frac{a^2}{b^2}\right),
\end{eqnarray}
and that in the same limit
\begin{eqnarray}
\eta(\tau\neq 0; a, b)&&
{\propto}
\frac{ a^2 - b^2 }{ \gamma^2}{\rightarrow} 0 .
\end{eqnarray}

\section{The gauge transformed amplitudes} 
\label{z/t}

In section III, the transition amplitudes $S_k^{(j)}$ and $A_k^{(j)}$ were 
defined and an exact relation, Eq.~(\ref{AS}), 
was established between them. However, this 
relation is not always useful as it involves infinite integrals and sums 
over the complete plane-wave basis set. In this appendix, 
a series expansion in inverse powers of the asymptotic time 
will be shown to relate $S_k^{(j)}$ and $A_k^{(j)}$ in a simpler 
way for symmetric collisions and calculations in the collider frame.

Two evolution operators can be defined for $|\Phi\rangle$ and $|\Psi\rangle$,
from Eq.\ (\ref{dirac}) and (\ref{gt_dirac}), respectively,
   \begin{eqnarray}
&&|\Phi(\vec{r}, t_f)\rangle = \hat{\cal{V}} (t_f,t_i) |\Phi(\vec{r}, 
t_i)\rangle \\
&&|\Psi(\vec{r}, t_f)\rangle = \hat{\cal{U}} (t_f,t_i) |\Psi(\vec{r}, 
t_i)\rangle. \end{eqnarray}
They are related by the gauge transformation of Eq.~(\ref{newgauge})
   \begin{eqnarray}
\hat{\cal{U}} (t_f,t_i) = U(z,t_f) \hat{\cal{V}} (t_f,t_i) 
U^{\dagger}(z,t_i) . \end{eqnarray}
The amplitudes are given by
   \begin{eqnarray}
&&S_k^{(j)} = \lim_{\stackrel{t_i\rightarrow -\infty}{t_f\rightarrow 
\infty}} \langle \chi_k (\vec{r},t_f) |\hat{\cal{V}} (t_f,t_i) | \chi_j 
(\vec{r},t_i)\rangle , \\ 
&&A_k^{(j)} = 
\lim_{\stackrel{t_i\rightarrow -\infty}{t_f\rightarrow 
\infty}} \langle \chi_k (\vec{r},t_f) |\hat{\cal{U}} (t_f,t_i) | \chi_j 
(\vec{r},t_i)\rangle .
\end{eqnarray}
A direct substitution gives
   \begin{eqnarray}
&&S_k^{(j)} = \lim_{\stackrel{t_i\rightarrow -\infty}{t_f\rightarrow
\infty}} \langle \chi_k (\vec{r},t_f) | U^{\dagger}(z,t_f) 
\hat{\cal{U}} (t_f,t_i) U(z,t_i) | \chi_j (\vec{r},t_i)\rangle .
\nonumber \\ \label{sa}
\end{eqnarray}

The asymptotic expressions,
   \begin{eqnarray}
&&U(z,t_i)\stackrel{t_i\rightarrow - \infty }{\longrightarrow}
\frac{[2\gamma(|t_i|+z)/b]^{iZ_A\alpha}}{[2\gamma(|t_i|-z)/b]^{iZ_B\alpha}} ,
\\
&&U^{\dagger}(z,t_f)\stackrel{t_f\rightarrow \infty }{\longrightarrow} 
\frac{[2\gamma(t_f-z)/b]^{iZ_A\alpha}}{[2\gamma(t_f+z)/b]^{iZ_B\alpha}},
\end{eqnarray}
reduce, for symmetric collisions, ($Z\equiv Z_A=Z_B$), to power series in 
$z/t$
   \begin{eqnarray}
&&U(z,t_i)\approx 1 +i2Z\alpha \frac{z}{|t_i|}+ ... \\
&&U^{\dagger}(z,t_f)\approx 1 - i2Z\alpha \frac{z}{t_f}+ ... .
\end{eqnarray}
Substituting these power series in Eq.~(\ref{sa}) and integrating term 
by term one gets,
   \begin{eqnarray}
S_k^{(j)} &\approx& A_k^{(j)} \nonumber \\
&& + {i2Z\alpha} 
\lim_{\stackrel{t_i\rightarrow -\infty}{t_f\rightarrow 
\infty}} \langle \chi_k (\vec{r},t_f) |\hat{\cal{U}} (t_f,t_i) 
\frac{z}{|t_i|} | \chi_j (\vec{r},t_i)\rangle 
\nonumber \\
&& - {i2Z\alpha}
\lim_{\stackrel{t_i\rightarrow 
-\infty}{t_f\rightarrow
\infty}} \langle \chi_k (\vec{r},t_f) |\frac{z}{t_f}  \hat{\cal{U}} 
(t_f,t_i) | \chi_j (\vec{r},t_i)\rangle + \cdots
\label{series1}
         \end{eqnarray}
Using completeness, Eq.\ (\ref{series1}) can equivalently be
written as
   \begin{eqnarray}
S_k^{(j)} &\approx& A_k^{(j)} 
\nonumber \\
&& 
+ {i2Z\alpha} 
 \sum_l \left[
A_k^{(l)}
\lim_{ t_i\rightarrow -\infty } 
 \langle \chi_k (\vec{r},t_f) | \frac{z}{|t_i|} | \chi_j (\vec{r},t_i)\rangle 
\right.
\nonumber \\
&& 
\left.
- 
A_l^{(j)}
\lim_{ t_f\rightarrow \infty} 
 \langle \chi_k (\vec{r},t_f) |\frac{z}{t_f} | \chi_j (\vec{r},t_i)\rangle 
\right]
 + \cdots
\label{series2}
         \end{eqnarray}

It is clear that $S_k^{(j)}$ and $A_k^{(j)}$ are in general different. 
For non-symmetric collisions, the relation between them involves, for 
example, a highly oscillatory, $z$-dependent phase which 
explicitly depends on $\gamma$.
However, this phase cancels for the interesting case of
symmetric collisions.
The first-order corrections to $S_k^{(j)} \approx 
A_k^{(j)} $ decrease linearly with time and have a functional nature 
of polarization effects. Higher-order corrections are higher moments of 
$z/t$.

\section{The discontinuity at the light fronts}
\label{discont}

In this appendix, the discontinuities of the spinor wave function at the 
light fronts (at $\tau_+= 0$ and at $\tau_-= 0$, 
excluding only $\tau_+=\tau_-= 0$) are deduced from Eq.~(\ref{sharp}). 
Previous derivations of the discontinuity of a wave function due to an 
ultra-relativistic charge are reviewed. 

At one light front ($\tau_+= 0$, $\tau_-\neq 0$),
Eq.~(\ref{sharp}) for $|G_+\rangle$ reads,
   \begin{eqnarray}
i \partial_{\tau_+} |G_+\rangle =\hat{h}_0 |G_-\rangle + 
B(\vec{r}_{\perp}) \delta (\tau_+) |G_+\rangle \; .
\label{g+}
\end{eqnarray}
The $\delta$-function singularity renders $|G_+\rangle$ discontinuous at 
$\tau_+= 0$, as can be seen by integrating both hand sides of 
Eq.~(\ref{g+}) 
with respect to $\tau_+$ from $-\epsilon$ to $\epsilon$ and taking the 
limit $\epsilon\rightarrow 0$,
   \begin{eqnarray}
|G_+(\tau_+= 0^+)\rangle \neq |G_+(\tau_+= 0^-)\rangle .
\end{eqnarray}

An auxiliary bi-spinor can be defined by a piece-wise gauge transformation,
   \begin{eqnarray}
|\tilde{G}_+\rangle \equiv 
      \exp [i B(\vec{r}_{\perp}) \theta(\tau_+)] |G_+\rangle .
\end{eqnarray}
Direct substitution gives,
   \begin{eqnarray}
i \partial_{\tau_+} |\tilde{G}_+\rangle =
\exp [i B(\vec{r}_{\perp}) \theta(\tau_+) ]
\hat{h}_0 |G_-\rangle 
\label{sg+}
\end{eqnarray}
The auxiliary bi-spinor is continuous at $\tau_+=0$, as can be seen by 
operating on both sides of Eq.~(\ref{sg+}) with 
$\lim_{\epsilon\rightarrow 0}\int_{-\epsilon}^{\epsilon}d \tau_+$, obtaining
   \begin{eqnarray}
|\tilde{G}_+(\tau_+= 0^+)\rangle = |\tilde{G}_+(\tau_+= 0^-)\rangle .
\end{eqnarray}

The continuity of $|\tilde{G}_+\rangle $ at $\tau_+=0$ ($\tau_- \neq 0$),
implies the discontinuity of Eq.~(\ref{dGB}) for $|G_+\rangle$.
Likewise, a continuity of 
   \begin{eqnarray}
|\tilde{G}_-\rangle 
     \equiv \exp [i A(\vec{r}_{\perp}) \theta(\tau_-)] |G_-\rangle 
\end{eqnarray}
at $\tau_-=0$, ($\tau_+ \neq 0$,)
implies the discontinuity of Eq.~(\ref{dGA}) for $|G_-\rangle$.

This Heavyside step-function, space-dependent, phase discontinuity 
was previously obtained in Ref.~\cite{Baltz97}. 
In earlier work \cite{Ja92},
a gauge transformation was used to establish 
the fact that the electromagnetic field of a charge which is moving at 
the speed of light can be equivalently given by gauge potentials with a 
$\delta$-function singularity at the light front, or by gauge potentials 
with only a step-function discontinuity there. The wave function of a 
particle interacting with this 
field is discontinuous or continuous, depending on the gauge choice. 
We choose to work with such a
gauge that would give a sharp interaction and a discontinuous spinor wave
function, yet we have used here other gauges to find the
discontinuities in an explicit form.

\section{The momentum-transfer distribution}
\label{Q_Z}

When the singularities at the light fronts are crossed, the 
transverse momentum changes.
The distribution for this momentum change 
is given in section IV by Eq.~(\ref{F}),
   \begin{eqnarray}
&&Q_{Z}(\vec{\kappa},\vec{b})\equiv
\frac{1}{(2\pi)^2}\int d \vec{r}_{\perp}
\ e^{i \vec{r}_{\perp}\cdot\vec{\kappa}}
\left[\frac{(\vec{r}_{\perp}-\vec{b})^2}{b^2}\right]^{- i \alpha Z}.
\end{eqnarray}
As mentioned in section IV,
divergence of $ Q_{Z}(\vec{\kappa},\vec{b}) $ is
an artifact of the ultra-relativistic approximation used
in Eq.\ (\ref{W_A}) and (\ref{W_B}).
The integral over this distribution converges and is normalized to 1
   \begin{eqnarray}
\int d \vec{\kappa} Q_{Z}(\vec{\kappa},\vec{b}) = 1 .
\end{eqnarray}
For a vanishing charge,
$Q_{Z=0}(\vec{\kappa},\vec{b})=\delta(\vec{\kappa})$,
but for finite charge $Z\neq 0$, this distribution
diverges both for vanishing and finite momentum transfer
$\vec{\kappa}$. In this appendix, we show that for $\vec{\kappa}\neq 0$ in
the perturbative limit, $\alpha Z\ll 1$, and after proper
regularization, the leading order correction to the
$\delta$-function is given by Eq.~(\ref{pertF}), i.e.
\begin{eqnarray}
Q_{Z}(\vec{\kappa},\vec{b}) \rightarrow \delta(\vec{\kappa})
-\frac{i \alpha Z}{\pi}
\frac{1}{\kappa^2} \exp (i \vec{b}\cdot\vec{\kappa}).
\end{eqnarray}

Integrating first over the angular variable,
\begin{eqnarray}
Q_{Z}(\vec{\kappa}\neq 0,\vec{b})
=\frac{b^2 \exp(i \vec{b}\cdot\vec{\kappa})}{2\pi}
\int_0^{\infty} d s J_0[s b \kappa] s^{1-i 2 \alpha Z},
\end{eqnarray}
where $b=|\vec{b}|$, $\kappa=|\vec{\kappa}|$, and $J_0$ is the Bessel
function.
The integral over $s$ diverges, but can be regulated for finite
$\kappa$ in the limit of $\alpha Z\ll 1$.

Using, for example, Eq.~(6.631.1) in Ref.\cite{GR} and Eq.~(13.5.1) in
Ref.\cite{AS} one has
\begin{eqnarray}
&& \lim_{\epsilon\rightarrow 0}
\frac{1}{2\pi}\int_0^{\infty} d s e^{-\epsilon s^2}
J_0[s b \kappa] s^{1-i 2 \alpha Z} \nonumber \\
&& = \lim_{\epsilon\rightarrow 0}
\frac{\Gamma(1-i \alpha Z)}{4\pi \epsilon^{1-i \alpha Z}}
\ {}_1F_1\left(1-i \alpha Z,1; \frac{- (b \kappa)^2}{4\epsilon}\right)
\nonumber \\
&& = \lim_{\epsilon\rightarrow 0}
\frac{1}{4\pi\epsilon} \exp[-(b \kappa)^2/4\epsilon]
\left(\frac{i b \kappa}{2\epsilon}\right)^{- i 2\alpha Z}
\nonumber \\
&& \ \ - \frac{i\alpha Z}{\pi} (b \kappa)^{-2} e^{-\pi\alpha Z}
\frac{\Gamma(-i \alpha Z)}{\Gamma(+i \alpha Z)} \left(
\frac{i b \kappa}{2}\right)^{+ i 2\alpha Z} ,
\end{eqnarray}
where ${}_1F_1$ is the confluent hypergeometric function.
Eq.~(\ref{pertF}) is now obtained by taking $\alpha Z \rightarrow 0$ and
using $\delta(b \vec{\kappa})\equiv \lim_{\epsilon\rightarrow 0}
\frac{1}{4\pi\epsilon} \exp [{-(b \kappa)^2/4\epsilon}] $.
Note that the limit $\epsilon\rightarrow 0$ can only be taken {\it after}
taking the perturbative limit $\alpha Z \rightarrow 0$.

\section{The transition current} 
\label{current}

As we are unaware of an appropriate reference,
we prove in this appendix that the transition 4-current
density defined in Eq.\ (\ref{tcurrent}) is 
conserved.
In fact, any two solutions of the free Dirac equation can be used to define
a conserved current in a similar way.
This proof is very similar to the one found 
in textbooks proving the probability current
to be conserved \cite{greiner}. 

Both $\chi_k$ and $\psi^{(j)}$ solve in region IV the
free Dirac equation in the Dirac representation
   \begin{eqnarray}
i\frac{\partial}{\partial t} \psi^{(j)}(\vec{r},t) =
\left[ 
 -i \check{\alpha}\cdot\vec{\nabla} + \check{\gamma^0} 
 \right] 
\psi^{(j)}(\vec{r},t) \; ,
\label{fde1} \\
i\frac{\partial}{\partial t} \chi_k(\vec{r},t) =
\left[ 
 -i \check{\alpha}\cdot\vec{\nabla} + \check{\gamma^0} 
 \right] 
\chi_k(\vec{r},t) \; .
\label{fde2}
\end{eqnarray}
Multiplying Eq.\ (\ref{fde1}) from the left by 
the adjoint of $\chi_k$ gives
\begin{equation}
i \chi_k^\dagger \frac{\partial \psi^{(j)}}{\partial t} 
=
-i \chi^\dagger_k \check{\alpha}\cdot\vec{\nabla} \psi^{(j)}
+ \chi^\dagger_k \check{\gamma^0} \psi^{(j)} \; .
\label{current1}
\end{equation}
Multiplying the Hermitian conjugate of Eq.\ (\ref{fde2})
from the right by $\psi^{(j)}$ gives
\begin{equation}
- i \frac{\partial \chi_k^\dagger}{\partial t} \psi^{(j)}
=
 i \left( \check{\alpha}\cdot\vec{\nabla} \chi^\dagger_k \right) \psi^{(j)}
+ \chi^\dagger_k \check{\gamma^0} \psi^{(j)} \; .
\label{current2}
\end{equation}
Subtracting Eq.\ (\ref{current2}) from Eq.\ (\ref{current1}) gives
\begin{equation}
 \frac{\partial }{\partial t} \left( \chi_k^\dagger\psi^{(j)} \right)
=
- \vec{\nabla} \cdot  \left(\chi^\dagger_k  \check{\alpha} \psi^{(j)} 
\right) \;, \label{current3}
\end{equation}
where the Hermiticity of the Dirac matrices has been used.
Using the definition of the transition current in Eq.\ (\ref{tcurrent}),
Eq.\ (\ref{current3}) is reveled as the continuity equation
\begin{equation}
 \frac{\partial }{\partial t} J_0^{(j,k)}  
+ \vec{\nabla} \cdot  \vec{J}^{(j,k)}
=0 \; ,
\label{current4}
\end{equation}
proving the transition-current density to be conserved.

\section{Application of Gauss' theorem}
\label{gauss_app}

As appendix \ref{current} shows, the transition 4-current,
defined in Eq.\ (\ref{tcurrent}),
is a conserved quantity, 
\begin{equation}
 \frac{  \partial J^\mu  } {  \partial x^\mu  }  =0  .
\label{4div}
\end{equation}
Integrating Eq.\ (\ref{4div}) over any empty space-time
hyper-volume, $V$, and applying Guass' theorem to convert
the volume integral into a surface integral over
the hyper-surface $S$ enclosing $V$, in general gives,
\begin{equation}
\int_S d \sigma J^\mu n_\mu = 0 \;,
\label{surface_int}
\end{equation}
where the unit $4$-vector $n_\mu$ is defined as the outward
pointing normal to $S$.

For our purposes, it is useful to apply Eq.\ (\ref{surface_int})
to the space-time region IV, defined in Fig.\ \ref{light_fronts}
by $\tau_{\pm} > 0$.
The closed hyper-surface $S$ enclosing region IV is made of the 
following open hyper-surfaces:
 (i)  $ t=t_f \rightarrow + \infty $, 
(ii) $ \tau_+ = 0^+$, $\tau_- > 0$, 
(iii) $ \tau_- = 0^+$, $\tau_+ > 0$, 
(iv) $ x \rightarrow \pm \infty $, and
(v) $ y \rightarrow \pm \infty $; (see Fig.\ \ref{gauss_surf2}).
Writing Eq.\ (\ref{surface_int}) for this surface gives
\begin{eqnarray}
0 &=&
  \lim_{t_f \rightarrow \infty} 
   \int d r_\perp \int_{+\infty}^{-\infty} J_0(\vec{r},t_f) \nonumber \\
&& - 2 \int d r_\perp 
   \int_{+\infty}^{0^+} d\tau_- J_+(\vec{r}_\perp, \tau_+=0^+, \tau_- )
  \nonumber \\
&& - 2 \int d r_\perp 
   \int^{+\infty}_{0^+} d\tau_+ J_-(\vec{r}_\perp, \tau_+, \tau_-=0^+ ) ,
\label{3sides}
\end{eqnarray}
where use was made of the fact that in any physical situation, i.e. for
a square-integrable wavepacket, the currents vanish as 
$ \vec{r} \rightarrow \infty$. The hyper-surfaces (iv) and (v) do not 
contribute to the integral of Eq.\ (\ref{surface_int}).
The factors of $2$ arise from the Jacobian
relating the original differentials to the differentials
for the light-front variables,
and the negative sign in the second and third terms arise
because the unit normal vectors $ \hat{n}_\pm $ are directed
outside the volume $V$, 
i.e.\ $ J \cdot \hat{n}_\pm = - J_\pm $. 
This completes our proof of Eq.\ (\ref{gauss}).




\pagebreak

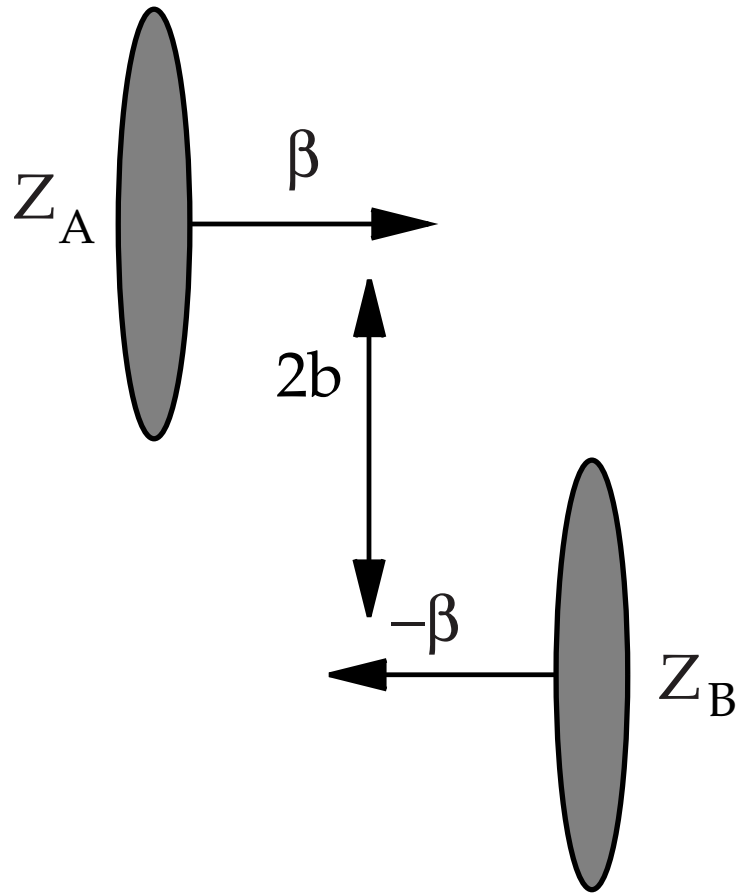
\begin{figure}[h]
\caption{
Schematic diagram depicting a relativistic heavy-ion collision
of two charges, $Z_A$ and $Z_B$, in the center-of-velocity frame
with impact parameter $2b$ and velocity $\beta$.
Lorentz contraction is extreme, so the ions are depicted as
oblate spheroids.
}
\label{coll_fig}
\end{figure}

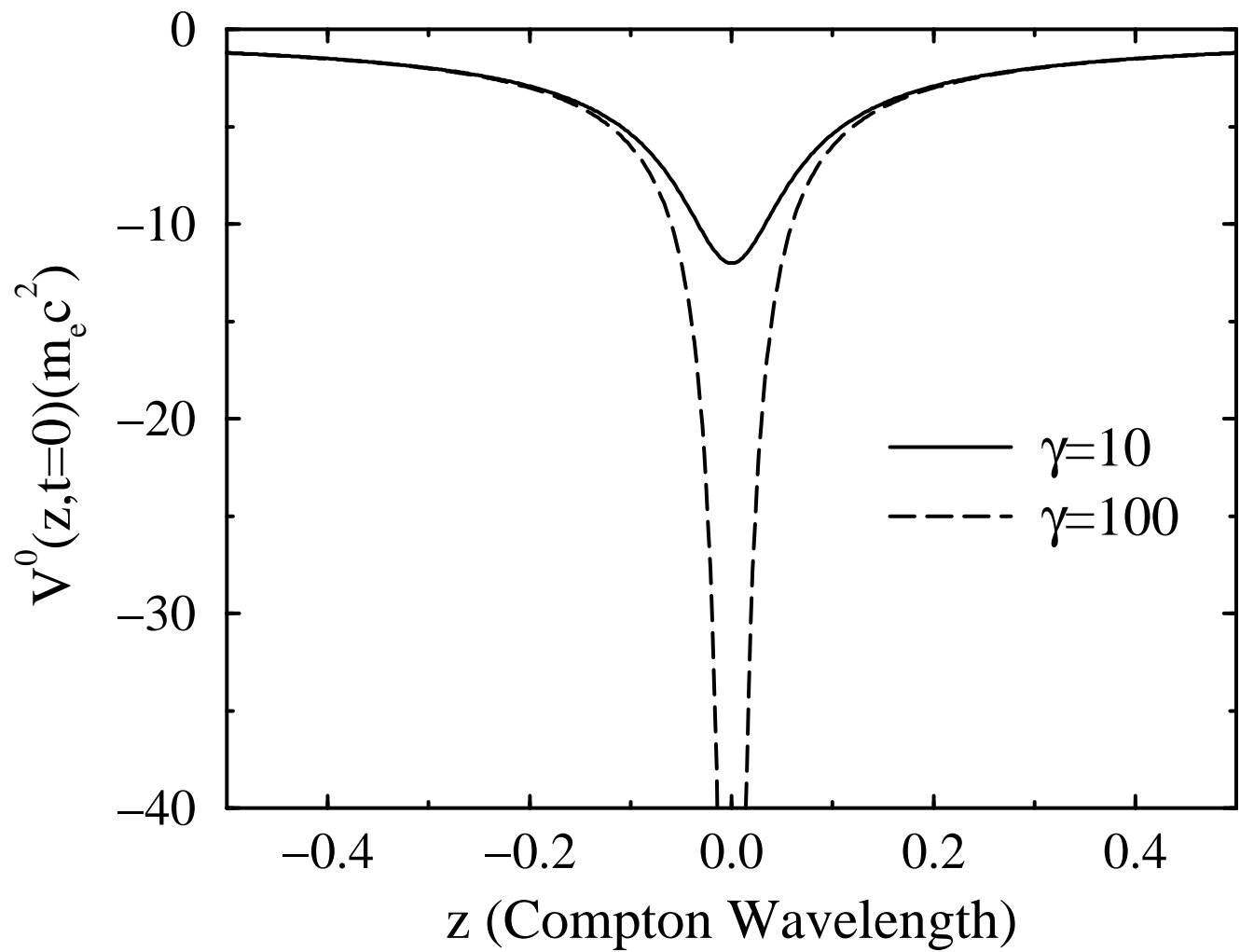
\begin{figure}[h]
\caption{
Shown is the scalar component of the Lorentz-gauge interaction, $V^0$,
for two different energies, 
(a) $\gamma=10$ (CERN-SPS energies), 
and (b) $\gamma=100$ (RHIC energies),
plotted as a function of a narrow range of the $z$-coordinate
for $t=0$, $\vec{b}=(1,0)$, and $\vec{r}_\perp=(2,0)$.
Notice that away from the vicinity of $z=0$ this interaction
is insensitive to the change in the energy of the ion.
}
\label{lorentz_int}
\end{figure}

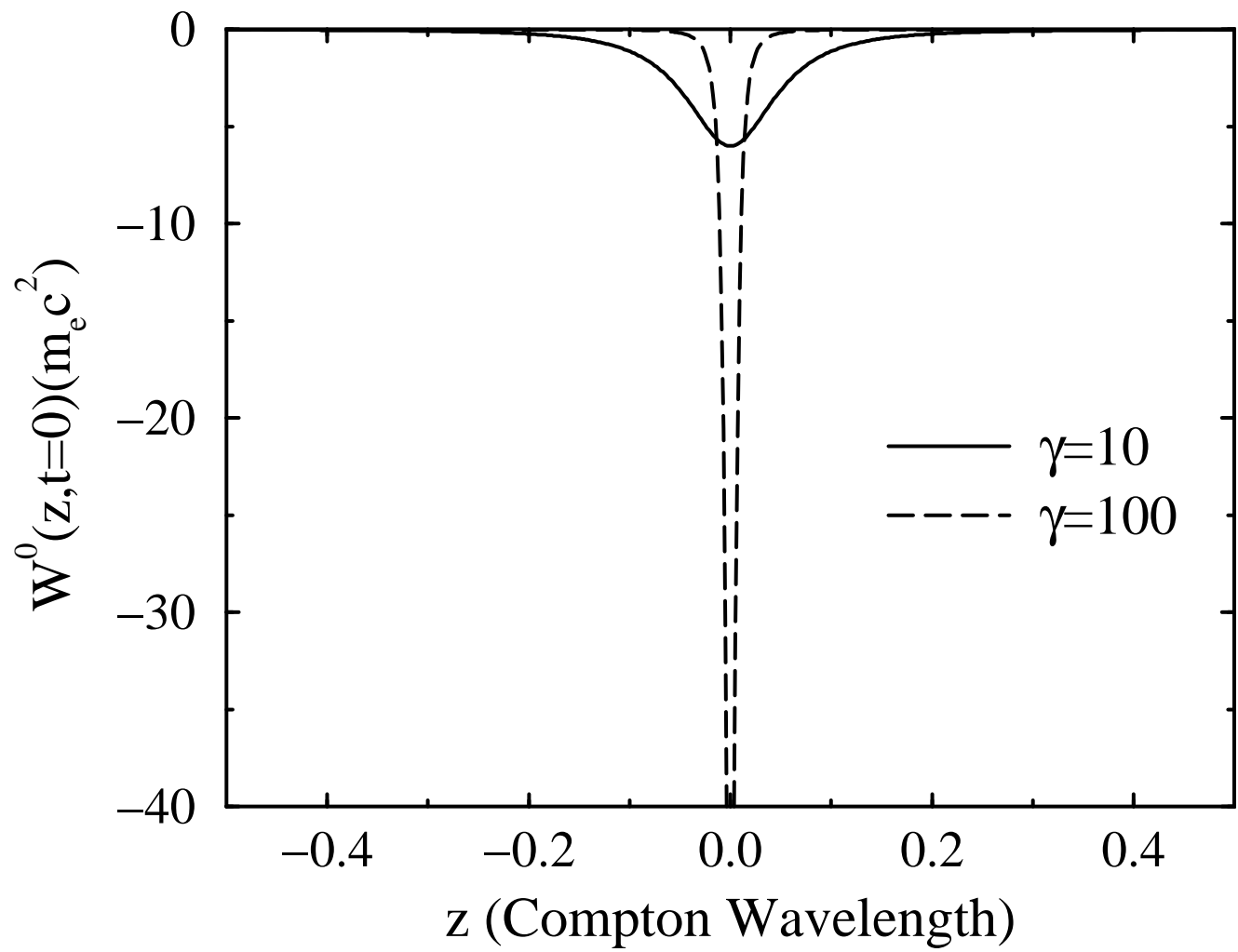
\begin{figure}[h]
\caption{
Same as Fig.\ \protect\ref{lorentz_int}\protect , 
except that here the scalar component of the
gauge-transformed interaction, $W^0$, is plotted.
Notice that the gauge-transformed interaction is short-ranged,
and that the range of the interaction decreases as the ion's
energy increases.
}
\label{eichler_int}
\end{figure}

\begin{figure}[h]
\caption{
Schematic diagram of the time-history of a heavy-ion collision
in the ultra-relativistic limit.
Motion in the $t-z$ plane is shown with $\vec{r}_\perp$, and
thus $\vec{b}$, assumed orthogonal to this plane.
Ions $A$ and $B$ move toward each other in the $z$ direction
with velocity $\beta = 1$.
The dotted lines are the projections of the ions 
trajectories on the $t-z$ plane, which for $\beta = 1$ coincide with the 
intersection of this plane with the light fronts at $z= \pm t$.
}
\label{lightfronts_def}
\end{figure}

\begin{figure}
\caption{
The light fronts, i.e.\ the hypersurfaces defined by $\tau_{\pm} =0$,
divide space-time into $4$ distinct regions:
(I) $\tau_+ < 0$, $\tau_- < 0$;
(II) $\tau_+ > 0$, $\tau_- < 0$;
(III) $\tau_+ < 0$, $\tau_- > 0$;
(IV) $\tau_+ > 0$, $\tau_- > 0$.
}
\label{light_fronts}
\end{figure}

\begin{figure}
\caption{
Depicted is the intersection of the surface $S$
enclosing region IV (shown as the broken line)
with the $ \tau_+ - \tau_- $ plane.
Light-front components of the transition current, $ J_\pm$,
are shown flowing into region IV at the light fronts,
and the time-like component of the transition current, $J_0$,
is shown flowing out of IV at the constant, large time $t_f$.
}
\label{gauss_surf2}
\end{figure}

\begin{figure}
\caption{
On the left are two terms of our result Eq.\ (\protect\ref{amp}\protect) 
for the amplitudes,
indicated by their respective space-time maps:
I$\rightarrow$ II$\rightarrow$ IV, and I$\rightarrow$ III$\rightarrow$ IV.
On the right are shown the two Feynman diagrams of second-order 
perturbation theory\protect\cite{BS89a}\protect  with 
their respective time ordering.
Our result assumes large $ \gamma$.
Reference\protect\cite{BS89a}\protect  assumes 
small $ \alpha Z$.
Exact agreement between the two results
is obtained in the combined limit.
}
\label{figure7}
\end{figure}

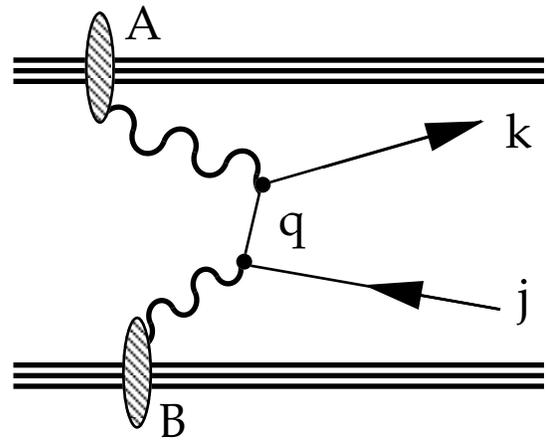
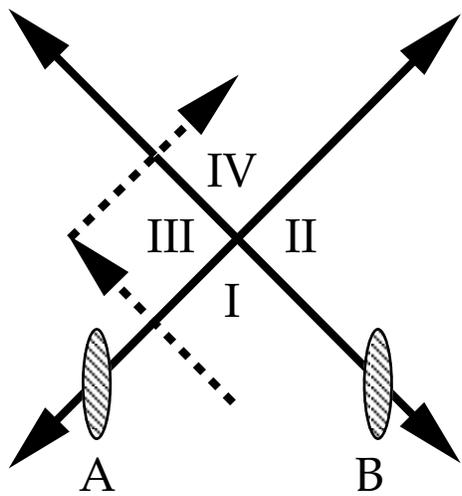
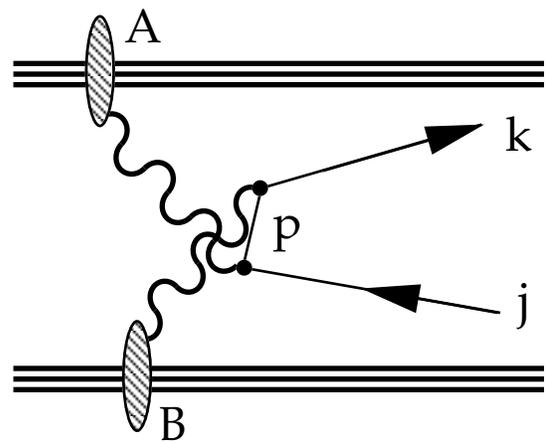
\end{document}